\documentclass[a4paper,11pt]{article}
\usepackage{pos}
\usepackage{braket}
\usepackage{subfig}
\usepackage{hyperref}

\begin{document}

\title{Numerical Stochastic Perturbation Theory around instantons}

\author*[a]{Paolo Baglioni}
\author[a]{Francesco Di Renzo}
\affiliation[a]{Dipartimento di Scienze Matematiche, Fisiche e Informatiche, Universit\`a di Parma \\and INFN, Gruppo Collegato di Parma, I-43124 Parma, Italy}
\emailAdd{paolo.baglioni@unipr.it}
\emailAdd{francesco.direnzo@unipr.it}
\abstract{Numerical Stochastic Perturbation Theory (NSPT) has over the years proved to be a valuable tool, in particular being able to reach unprecedented orders for Lattice Gauge Theories, whose perturbative expansions are notoriously cumbersome. One of the key features of the method is the possibility to expand around non-trivial vacua. 
While this idea has been around for a while, and it has been implemented in the case of the (non-trivial) background of the Schr\"odinger functional, NSPT expansions around instantons have not yet been fully worked out. Here we present computations for the double well potential in quantum mechanics. We compute a few orders of the expansion of the ground-state energy splitting in the one-instanton sector. We discuss how (already) known two-loop results are reproduced and present the current status of higher-order computations.}
\FullConference{The 39th International Symposium on Lattice Field Theory - LATTICE 2022\\
8th-13th August, 2022,\\
Rheinische Friedrich-Wilhelms-Universit\"at Bonn, Bonn, Germany}
\maketitle

\section{Non-perturbative physics from instantons}

Since their first introduction, instantons have been shown over the years to be of fundamental importance for the complete understanding of certain physical phenomena. 
Instantons are classical solutions of the euclidean equations of motion and mediate barrier-penetration processes, often assumed to play a fundamental role in determining the ground-state structure of theories such as QCD \cite{schafer_instantons_1998}.
In particular they are classical configurations for which the action is finite and usually give rise to non-perturbative processes.\\
There is a plethora of models that display the presence of instantons, both in non-relativistic quantum mechanics and in quantum field theory \cite{rajaraman_solitons_1982}. Among the former the most paradigmatic is the double well potential
\begin{equation}
	V(x) = \lambda(x^2-x_0^2)^2.
\end{equation}
In this case the presence of instantons solves the degeneracy problem for the energy levels introducing a non-perturbative splitting proportional to $e^{-A/\lambda}$ (with $A>0$) and thus invisible with standard perturbation theory. Neglecting multi-instanton contributions, the ground-state energy can be understood as a sum of a perturbative series over the trivial vacuum and a perturbative series in the one-instanton sector: 
\begin{equation}
	E_0(\lambda) = \sum_{n=0}^{\infty}\lambda^n E_{0,0}^{(n)} + e^{-\frac{A}{\lambda}}\sum_{n=0}^{\infty}\lambda^n E_{0,1}^{(n)} + \dots
\end{equation}
Though this system has been studied extensively in the past decades, the main attempts to compute perturbative corrections on top of the instantons are based on the use of WKB techniques on the one hand and Path Integrals (PI) on the other. Modern WKB appears manageable and powerful in simple systems while PI formalism  seems more suitable for generalization to higher-dimensional theory; on the other side in this approach perturbative computations become hard already at not-so-high orders. Having in mind as the ultimate goal the study of QCD, in this work we will focus only on the PI formalism. 

\subsection{Computing perturbative corrections in the one-instanton sector}

Computing the energy splitting coefficients for the ground-state is generally a cumbersome task: technicalities make these calculations challenging even at second-order. In the double well potential case this is especially true. One takes into account parity-reflection symmetry and introduces an additional quantum number $\pm$ for indexing the eigenvalues; in particular, for the ground state energies we have
\begin{equation}
    	E_{0,\pm} = E_{0} \mp \frac{\Delta E_0}{2}.
\end{equation}
The energy splitting contribution is proportional to $e^{-\frac{A}{\lambda}}$. The euclidean partition function for small coupling constant $\lambda$ and large $\beta$ reads
\begin{equation}
	\lim_{\beta\to\infty} Z(\beta) = \lim_{\beta\to\infty} \int dx\ \braket{ x|e^{-\beta \hat{H}}|x} \approx e^{-\beta E_{0,+}} + e^{-\beta E_{0,-}} \approx 2e^{-\frac{\beta}{2}(E_{0,+} + E_{0,-})}\cosh{\frac{\beta\Delta E_0}{2}}.
\end{equation}
This quantity seems inappropriate to lift the level's degeneracy since it is dominated by the purely perturbative sector. In contrast the twisted partition function exhibits a non-vanishing contribution coming from the one-instanton sectors
\begin{equation}
	\lim_{\beta\to\infty} Z_a(\beta) = \lim_{\beta\to\infty} \int dx\ \braket{ -x|e^{-\beta \hat{H}}|x} \approx e^{-\beta E_{0,+}} - e^{-\beta E_{0,-}} \approx 2e^{-\frac{\beta}{2}(E_{0,+} + E_{0,-})}\sinh{\frac{\beta\Delta E_0}{2}}.
\end{equation}
Thus the energy splitting can be extracted in perturbation theory from the ratio
\begin{equation}
\label{eq:ration}
	\lim_{\beta\to\infty} \frac{Z_a(\beta)}{Z(\beta)} = \frac{\beta \Delta E_0}{2} = e^{-\frac{A}{\lambda}} C \Bigl( 1 + \lambda z_1 + \lambda^2 z_2 + \dots\Bigr).
\end{equation}
In view of this one has to perturbatively compute the two partition functions in Eq.~(\ref{eq:ration}) and reorganize the ratio order-by-order. As usual, the first step is to look for the minimal action configuration for the twisted partition function: this is the non-trivial vacuum state with anti-periodic boundary conditions (ABC) one has to expand around. To compute perturbative expansions one then needs to write down the propagator, the vertices and to generate the Feynman diagrams. All in all this is a laborious task and until now only the first and second coefficients have been computed \cite{wohler_two-loop_1994, escobar-ruiz_three-loop_2015}.

\section{Numerical Stochastic Perturbation Theory around non-trivial vacua}

In this work we aim to take advantage of Numerical Stochastic Perturbation Theory (NSPT) by expanding the lattice theory around non-trivial solutions. 	NSPT \cite{di_renzo_weak_1995,di_renzo_four_1994,di_renzo_numerical_2004} can be seen as a numerical implementation of Stochastic Perturbation Theory, a theoretical framework formulated from the famous work on Stochastic Quantization \cite{parisi_perturbation_1981,damgaard_stochastic_1987}. For a recent application of NSPT in LGT see \cite{del_debbio_large-order_2018}. \\
Starting from the euclidean (lattice) action $S[x_i]$, we introduce an extra degree of freedom $\tau$  and write $x_i \rightarrow x_i(\tau)$. An evolution takes place in the stochastic time according to the Langevin equation
\begin{equation}
\label{eq:langevin}
	\frac{dx_i(\tau)}{d\tau} = - \frac{\partial S[x_i]}{\partial x_i(\tau)} + \eta_i(\tau).
\end{equation}
The last term is the Gaussian noise term properly normalized:
\begin{equation}
	\left\langle \eta_i(\tau)\right\rangle_\eta = 0 \quad \quad \left\langle \eta_i(\tau)\eta_j(\tau')\right\rangle_\eta = 2 \delta_{ij}\delta(\tau-\tau')
\end{equation}
and
\begin{equation}
\left\langle ... \right\rangle_\eta = \frac{\int D\eta_i(\tau) \ ... \ e^{-\frac{1}{4}\sum_j\int d\tau \eta_j(\tau)^2} }{\int D\eta_i(\tau)e^{-\frac{1}{4}\sum_j\int d\tau \ \eta_j(\tau)^2}}.
\end{equation}
The fundamental assertion of Stochastic Quantization is that \cite{parisi_perturbation_1981}
\begin{equation}
	\lim_{\tau\to\infty} \left\langle O[x_i(\tau) \dots x_k(\tau)] \right\rangle_\eta = \left\langle O[x_i \dots x_k] \right\rangle
\end{equation}
that is, in the limit of large stochastic time the expectation value of an observable with respect to Gaussian noise consistently reproduces the expectation value calculated in the PI formalism. For small coupling constant the fields can be expanded as power series of $\lambda$
\begin{equation}
\label{eq:pt}
	x_i(\tau) = x_i^{(0)}(\tau) + \sum_{n>0}\lambda^n x_i^{(n)}(\tau).
\end{equation}
Replacing this expression for the fields in the Langevin equation, the latter can be regarded as a tower of perturbative equations that are exact at any truncation order. One numerically integrates (with a chosen integrator) Eq.~(\ref{eq:langevin}) order-by-order. The perturbative coefficients of observables are in turn obtained by computing
\begin{equation}
	\left\langle O[x_i](\tau) \right\rangle_\eta = \left\langle O[\sum_n\lambda^n x_i^{(n)}(\tau)]\right\rangle_{\eta} = \sum_{n\ge 0}\lambda^n O^{(n)}(\tau).
\end{equation}
From a practical point of view, averages over Gaussian noise are traded for time averages over long Monte-Carlo history.
\subsection{Anti-periodic boundary conditions, zero modes and all that}
\begin{figure}
\centering
   {\includegraphics[width=.48\textwidth]{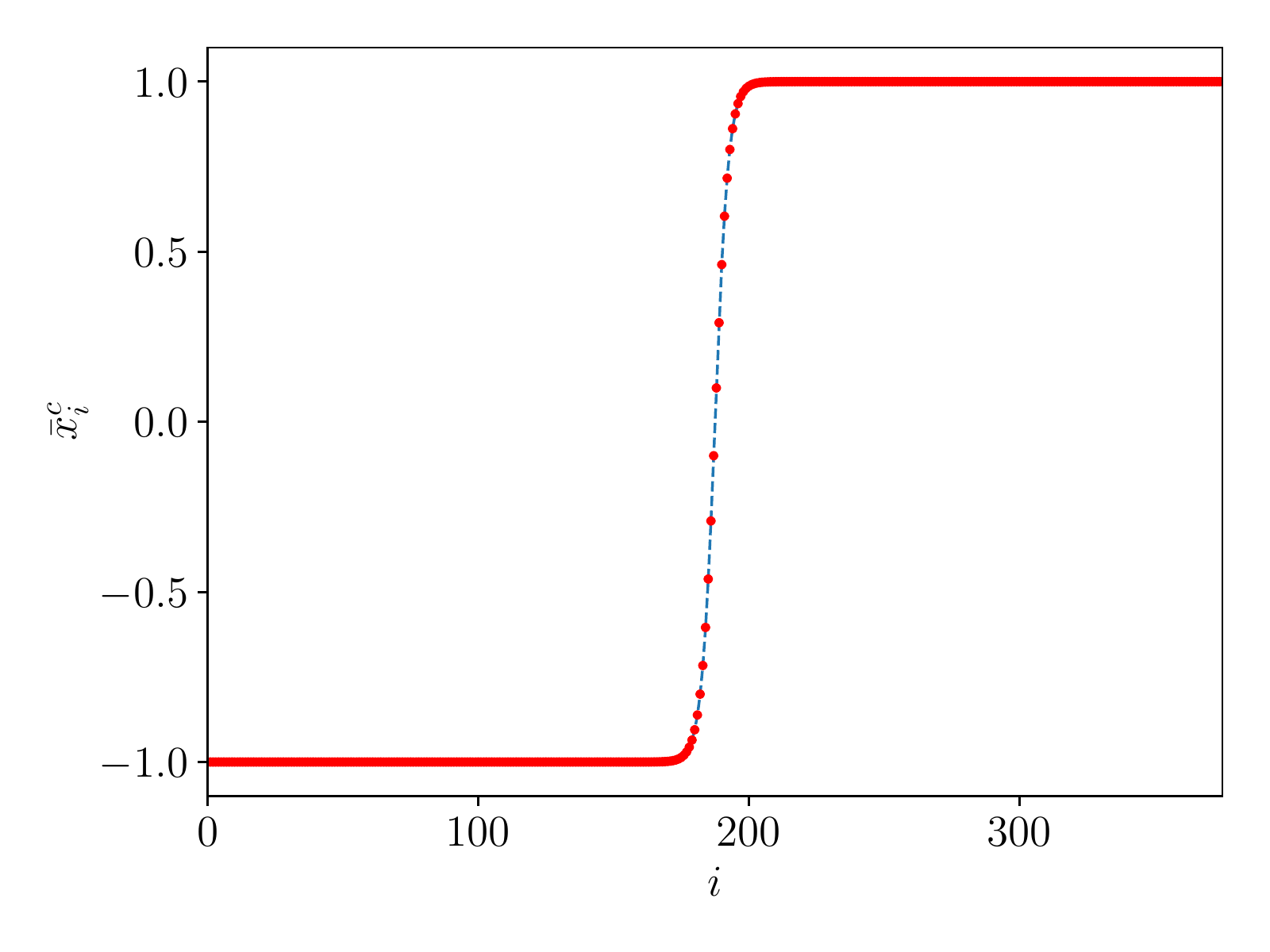}} \quad
   {\includegraphics[width=.482\textwidth]{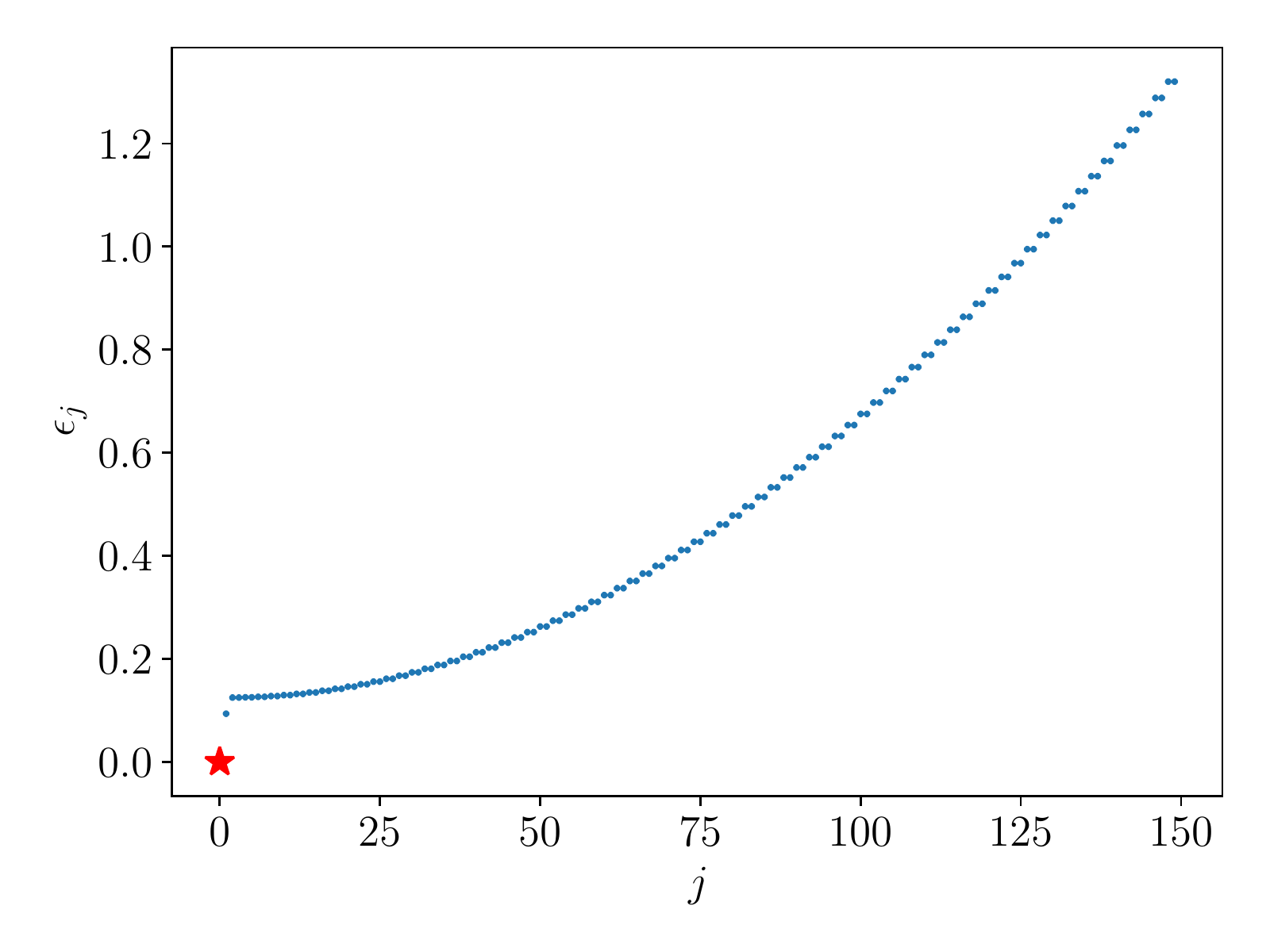}}
\caption{Left: Instantonic lattice solutions (red dots) compared with the instantonic solution in the continuum (blue dashed line). Right: eigenvalues of the kinetic operator in Eq.~(\ref{eq:kin_op}) (zero-mode emphasized in red).}
\label{fig:fig_1}
\end{figure}
Consider the lattice theory
\begin{equation}
	S_E[\tilde{x}_i] = \sum_{i=0}^{L} \Bigl[ \frac{1}{2}\tilde{m}(\tilde{x}_{i+1}-\tilde{x}_i)^2 + \tilde{\lambda}(\tilde{x}_i^2-\tilde{x}_0^2)^2\Bigr]
\end{equation}
written in terms of the adimensional parameters $\tilde{m}=ma$, $\tilde{x}_i=x_i/a$, $\tilde{\lambda}= \lambda a^5$ and $\tilde{x}_0=x_0/a$, where $a$ is the lattice spacing. 
$L$ has to do with infrared cut-off $T=La$ which unavoidably has to be there in a lattice simulation. The value $T=150$ has been chosen and verified to be largely subdominant with respect to finite $a$ effects. In view of this, any reference to it will be omitted in the following ({\em i.e.} we will always write $\sum_i$). Notice that this also means that no infinite volume limit will be taken.
It is trivial to find solutions of the classical equations of motion on the lattice with periodic boundary conditions (PBC): these are nothing but the constant field configurations $\tilde{x}_i = \pm \tilde{x}_0$. This does not hold true for anti-periodic boundary conditions. The classical solution is found by the steepest descent method, i.e. one looks for stationary solutions of
\begin{equation}
	\dot{\tilde{x}}_i = - \frac{\partial S_E}{\partial \tilde{x}_i} 
\end{equation}
where ABC imply $\tilde{x}_{L+1} = -\tilde{x}_0$ and $\tilde{x}_{-1} = - \tilde{x}_{L}$. This procedure returns the instantonic profile $\tilde{x}^*_i = \tilde{x}_0\cdot \bar{x}_i$ on the lattice (we give an example of this in Fig.~\ref{fig:fig_1}). The lattice theory for the quantum fluctuations $\tilde{\xi}_i = \tilde{x}_i - \tilde{x}_0\bar{x}_i$ reads
\begin{equation}
\label{eq:s}
	S_E = S_E[\tilde{x}^*] + \sum_i \Bigl[ \frac{1}{2} \tilde{m}(\tilde{\xi}_{i+1} - \tilde{\xi}_i)^2 + \frac{1}{2} \tilde{m}\tilde{\omega}^2\Bigl( \frac{3}{2}\bar{x}_i^2 - \frac{1}{2}\Bigr)\tilde{\xi}_i^2 + \sqrt{2\tilde{\lambda}\tilde{m}\tilde{\omega}}\bar{x}_i\tilde{\xi}_i^3 + \tilde{\lambda}\tilde{\xi}_i^4   \Bigr]
\end{equation}
where $4\tilde{\lambda}\tilde{x}_0^2 = \frac{1}{2}\tilde{m}\tilde{\omega}^2$ to lead us back to the free theory in \cite{escobar-ruiz_three-loop_2015}.
    In analogy with the continuum theory \cite{zinn-justin_path_2004}, the kinetic operator 
\begin{equation}
\label{eq:kin_op}
	K_{ij} = \frac{\partial^2 S_E[\tilde{\xi}]}{\partial \tilde{\xi}_i \partial \tilde{\xi}_j}\bigg|_{\tilde{\xi}=0}
\end{equation}
was found to have a vanishing eigenvalue (see Fig.~\ref{fig:fig_1}). This means that in the (generalized) momentum-space the zero-mode has no damping force and can propagate freely along the Monte-Carlo history, compromising numerical stability.
\begin{figure}
\centering
   {\includegraphics[width=.482\textwidth]{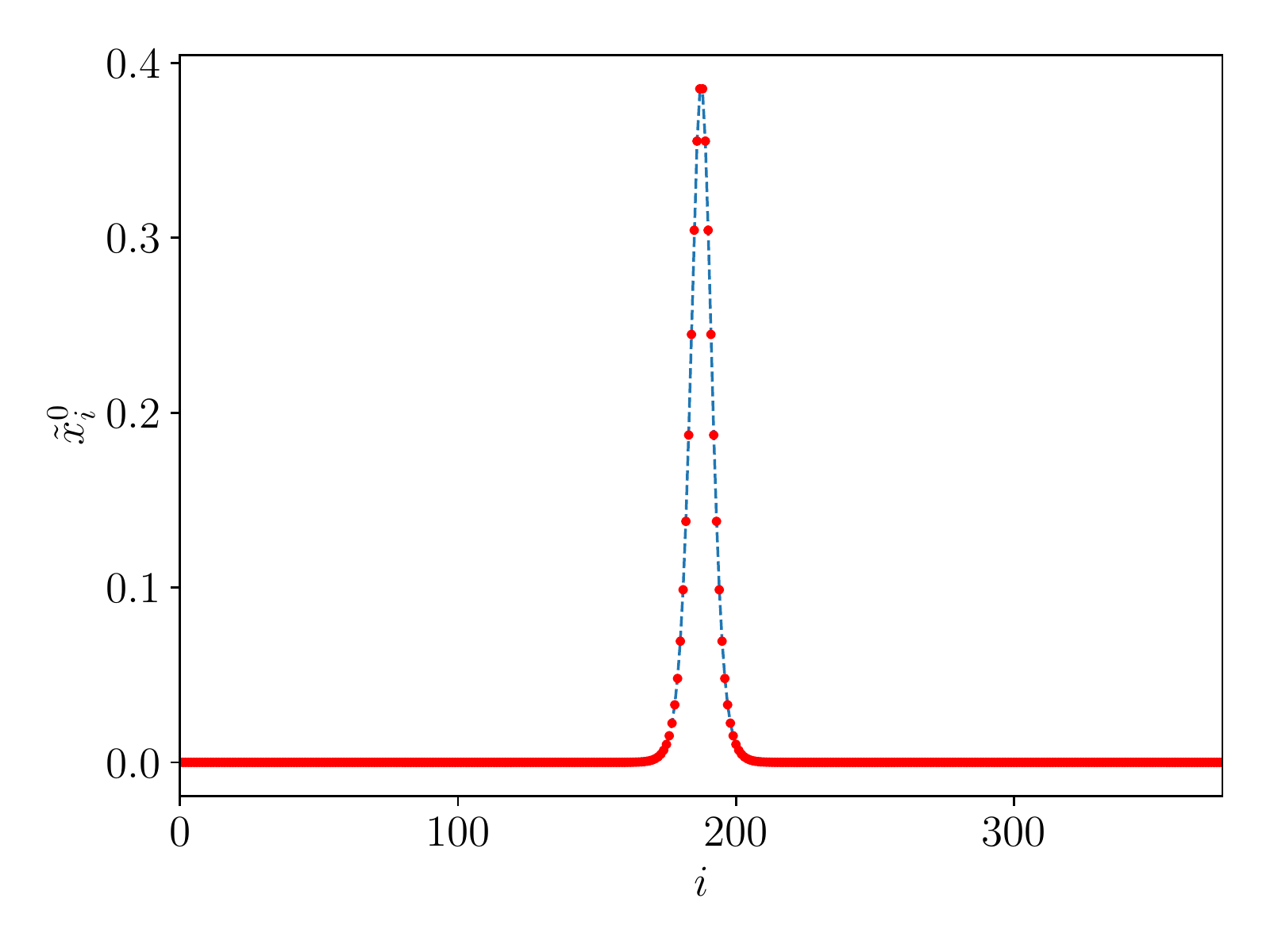}}
\caption{Zero-mode profile for the lattice theory in Eq.~(\ref{eq:s}) (red dots) and the continuous one (blue dashed line).}
\label{fig:fig_2}
\end{figure}
Tracking the correspondence with the continuum theory \cite{jentschura_multi-instantons_2011} and in order to evolve only orthogonal components to the zero-mode, the following decomposition was introduced
\begin{equation}
\label{eq:decomposition}
	\tilde{\xi}_i = c_0 \tilde{x}^0_i + \tilde{\xi}_i^\perp  \rightarrow \tilde{\xi}_i^\perp = \tilde{\xi}_i - c_0 \tilde{x}^0_i \quad \quad c_0 = \sum_i \tilde{\xi}_i \tilde{x}_i^0
\end{equation}
where $\tilde{x}^0_i$ is the eigenvector of the operator defined in Eq.~(\ref{eq:kin_op}) corresponding to a vanishing eigenvalue (we show the shape of the lattice zero-mode profile in Fig.~\ref{fig:fig_2}). We will now aim at expressing the partition function in terms of $\tilde{\xi}^\perp_i$. It should be noted that, in view of Eq.~(\ref{eq:pt}) , Eq.~(\ref{eq:decomposition}) is to be understood order-by-order.
The twisted partition function
\begin{equation}
	\label{eq:Za}
	Z_a = \int_{ABC} \prod_i d\tilde{x}_i e^{-S_E[\tilde{x}]} = e^{-S_E[\tilde{x}^*]}\int_{ABC} \prod_i d\tilde{\xi}_i e^{-S_E[\tilde{\xi}]}
\end{equation}
can be regularized using the Faddeev-Popov method. Again, the lattice implementation takes inspiration from the continuous counterpart \cite{aleinikov_instantons_1987}. This amounts to writing a convenient representation of the identity
\begin{equation}
		1 = \int d\tau_0 \delta\Bigl( \sum_k(\tilde{x}_k - \tilde{x}^*_k(\tau_0))\tilde{x}_k^0(\tau_0)\Bigr)\Bigr[-\sum_k \dot{\tilde{x}}^*_k(\tau_0)\tilde{x}_k^0(\tau_0) + \sum_k(\tilde{x}_k - \tilde{x}^*_k(\tau_0))\dot{\tilde{x}}_k^0(\tau_0)\Bigl]
\end{equation}
in which $\tau_0$ parametrizes the family of instantonic solutions and can be interpreted as the tunneling time, which can occur at any point (time translation invariance).
Once the previous equation is inserted into Eq.~(\ref{eq:Za}) the Dirac delta enables to integrate out the zero-mode component. After a few algebraic steps the twisted lattice partition function reads
\begin{equation}
\label{eq:f_def}
	Z_a = \frac{ e^{-S_E[\tilde{x}^*]}\beta\sqrt{S}}{\sqrt{2\pi}} Z_a^\perp \left\langle[1 + \sqrt{\tilde{\lambda}} \sum_l\tilde{\xi}_{l}^\perp v_l \Bigr]\right\rangle_a^\perp = \frac{ e^{-S_E[\tilde{x}^*]}\beta\sqrt{S}}{\sqrt{2\pi}} Z_a^\perp (1+\tilde{\lambda}\tilde{f}^{(a)}_1 + \tilde{\lambda}^2\tilde{f}^{(a)}_2 + \dots )
\end{equation}
where $\braket{...}_a^\perp$ means the average over the anti-periodic theory without zero-mode, $Z_a^\perp$ is the corresponding twisted partition function and $v_l$ is a pure geometric profile given by
\begin{equation}
	v_l = \sqrt{\frac{8}{\tilde{m}\tilde{\omega}^2}}\frac{1}{a\bar{\gamma}}( \tilde{x}^0_{l-1} - \tilde{x}^0_{l}) \quad \mathrm{where} \quad \sqrt{\frac{\tilde{m}\tilde{\omega}^2}{8\tilde{\lambda}}}  \bar{\gamma} = \sum_l \tilde{x}^*_l \Bigl( \frac{\tilde{x}^0_{l-1} - \tilde{x}^0_{l}}{a} \Bigr) = \sqrt{S} .
\end{equation}
\begin{figure}
\centering
   {\includegraphics[width=.485\textwidth]{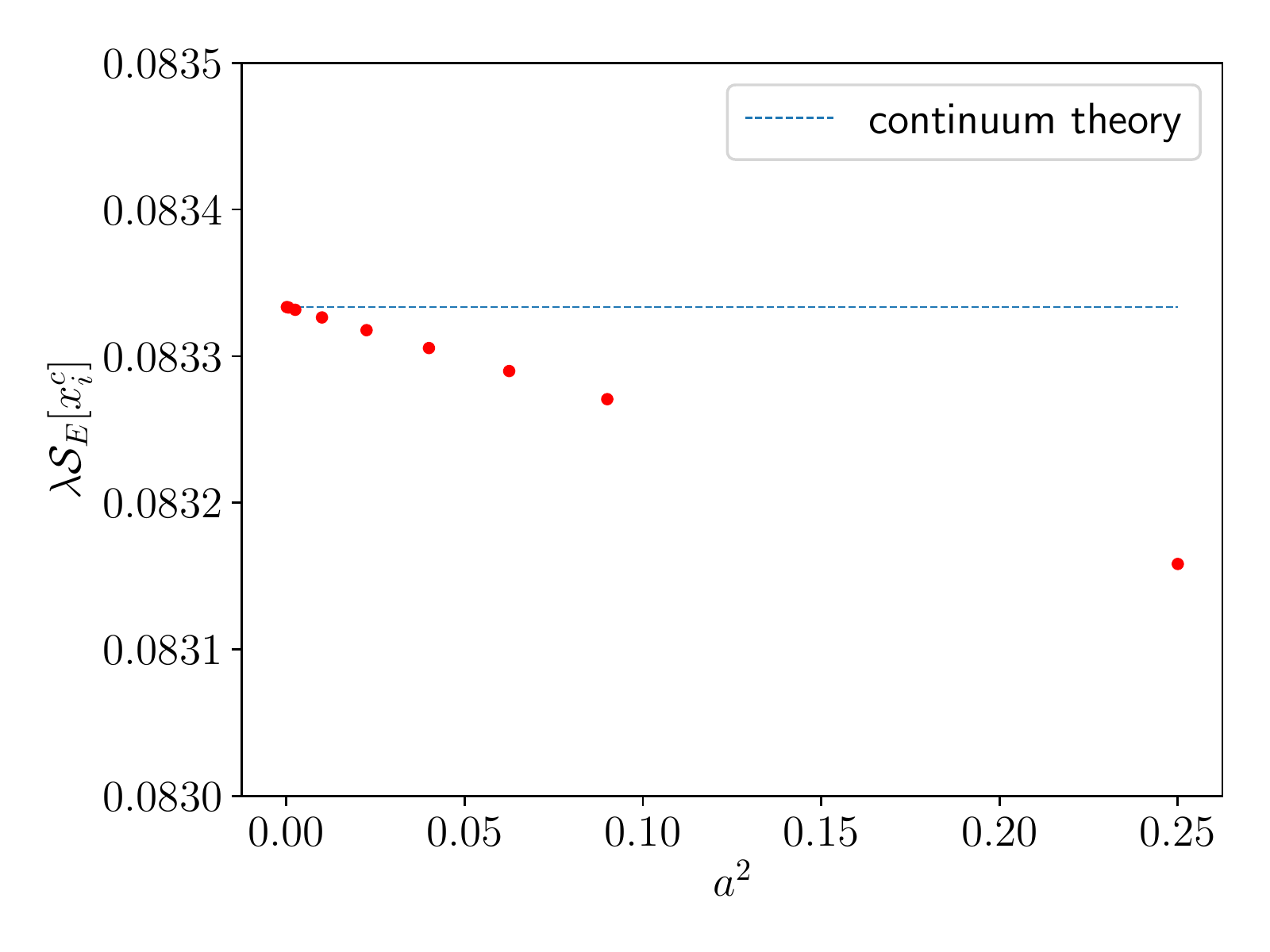}} \quad
   {\includegraphics[width=.48\textwidth]{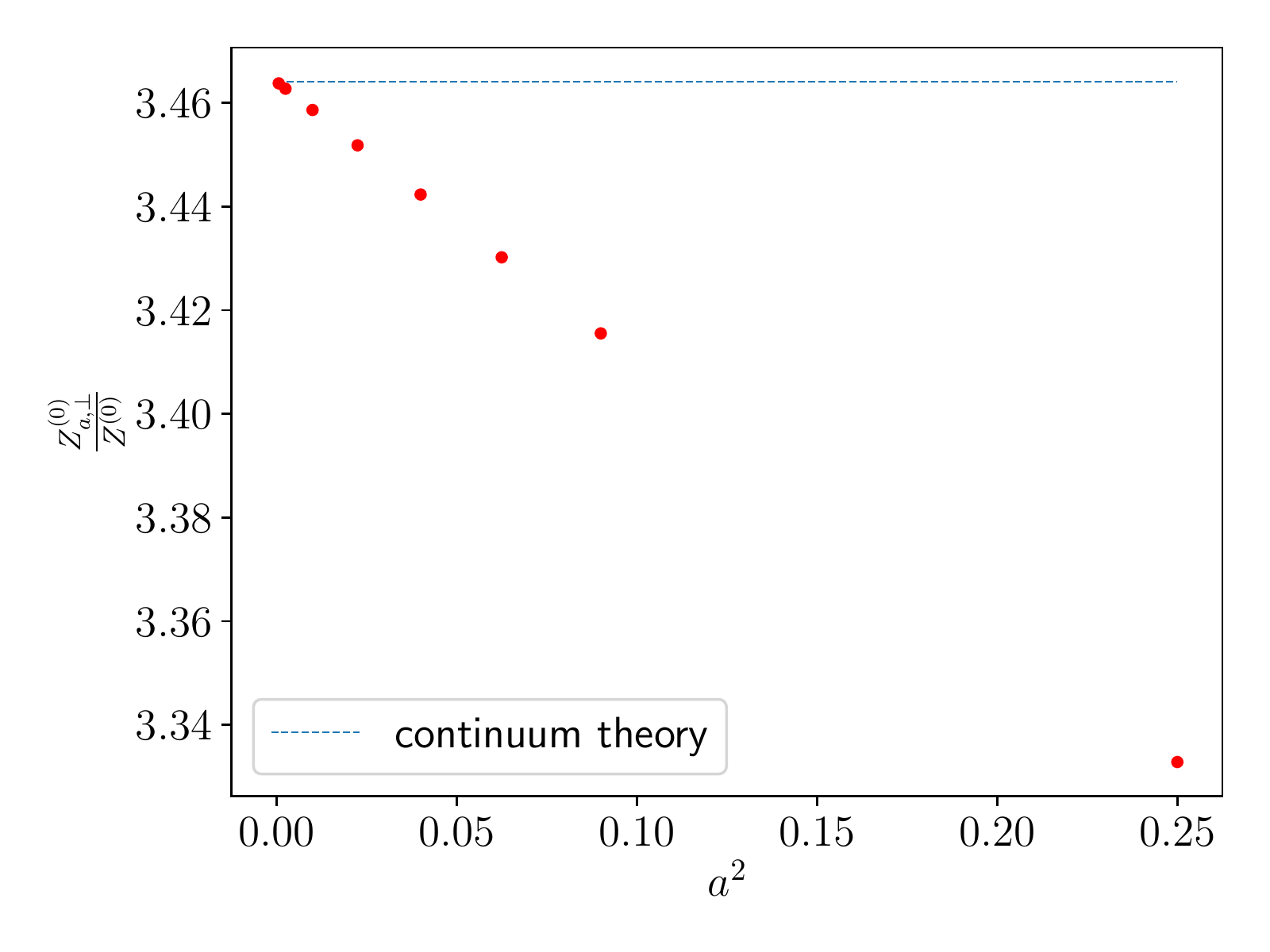}}
\caption{Numerical values for different lattice spacing of prefactors in Eq.~(\ref{eq:Za_over_Z}) (red points) and the continuum limit prediction (blue dashed lines).}
\label{fig:fig_3}
\end{figure}
The superscript $a$ in {\em e.g.} $\tilde{f}^{(a)}_1$ is there to remind that the theory is defined at a given lattice spacing $a$. Notice that the subscript $a$ ({\em e.g.} in $Z_a$) has a different meaning ({\em anti-periodic}). The notation is strictly speaking ambiguous, but we think the two different meanings are easy to recognize.\\ 
To compute the twisted partition function $Z_a^\perp$ we first compute
\begin{equation}
\label{eq:b_n}
	\left\langle  \frac{1}{2}\sqrt{2\tilde{\lambda}\tilde{m}\tilde{\omega}^2}\sum_i \bar{x}_i\tilde{\xi}_i^{\perp^3}+\tilde{\lambda}\sum_i \tilde{\xi}_i^{\perp^4} \right\rangle_a^\perp = \tilde{\lambda}\tilde{b}^{(a)}_1 + \tilde{\lambda}^2 \tilde{b}^{(a)}_2 + \tilde{\lambda}^3 \tilde{b}^{(a)}_3 \dots
\end{equation}
where the superscript $a$ has the same meaning as in Eq.~(\ref{eq:f_def}). 
$Z_a^\perp$ can be computed by exponentiation
\begin{equation}
\label{eq:z_a_p}
	Z_a^\perp = Z_{a,\perp}^{(0)}\cdot\exp{ \Bigl[ \sum_{n=1}^\infty \frac{\tilde{\lambda}^n}{n}\tilde{b}^{(a)}_n\Bigr] }.
\end{equation}
The prefactor $Z_{a,\perp}^{(0)}$ is nothing but the partition function for the free theory around the instantonic solution. The same is true for the theory with PBC.
In the end we find
\begin{equation}
\label{eq:Za_over_Z}
	\frac{Z_a}{Z} = \frac{e^{-S_E[\tilde{x}^*]}\beta\sqrt{S}}{\sqrt{2\pi}}\Bigl( \frac{Z_{a,\perp}^{(0)}}{Z^{(0)}} \Bigr)\left\langle[1 + \sqrt{\tilde{\lambda}} \sum_l\tilde{\xi}_{l}^\perp v_l \Bigr]\right\rangle_a^\perp \exp{\Bigl[ \sum_{n\ge1}\tilde{\lambda}^n\Bigl( \frac{\tilde{b}^{(a)}_n-\tilde{c}_n}{n} \Bigr) \Bigr]}
\end{equation}
where the terms $\tilde{c}_n$ are the counterpart of $\tilde{b}_n$ in Eq.~(\ref{eq:z_a_p}). Notice that in this work we take them from standard Quantum Mechanics perturbation theory computations (symbolic computations are performed in {\em Mathematica}) in the continuum limit, and thus have no superscript $a$ (for completeness we remind the reader that they do not have any infrared cut-off effect either). 

\section{Two-loop results}
\begin{figure}
\centering
   {\includegraphics[width=.48\textwidth]{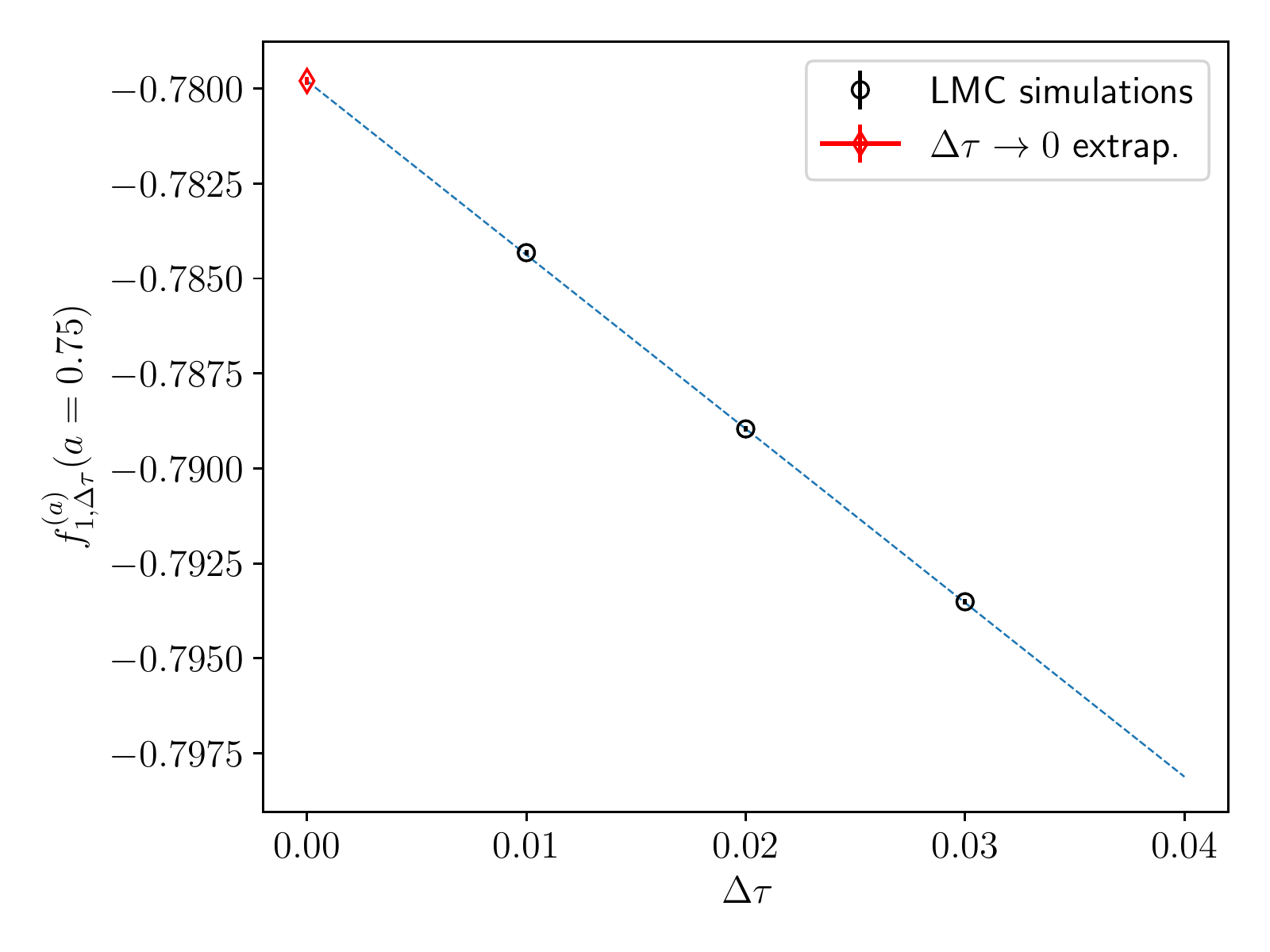}} \quad
   {\includegraphics[width=.482\textwidth]{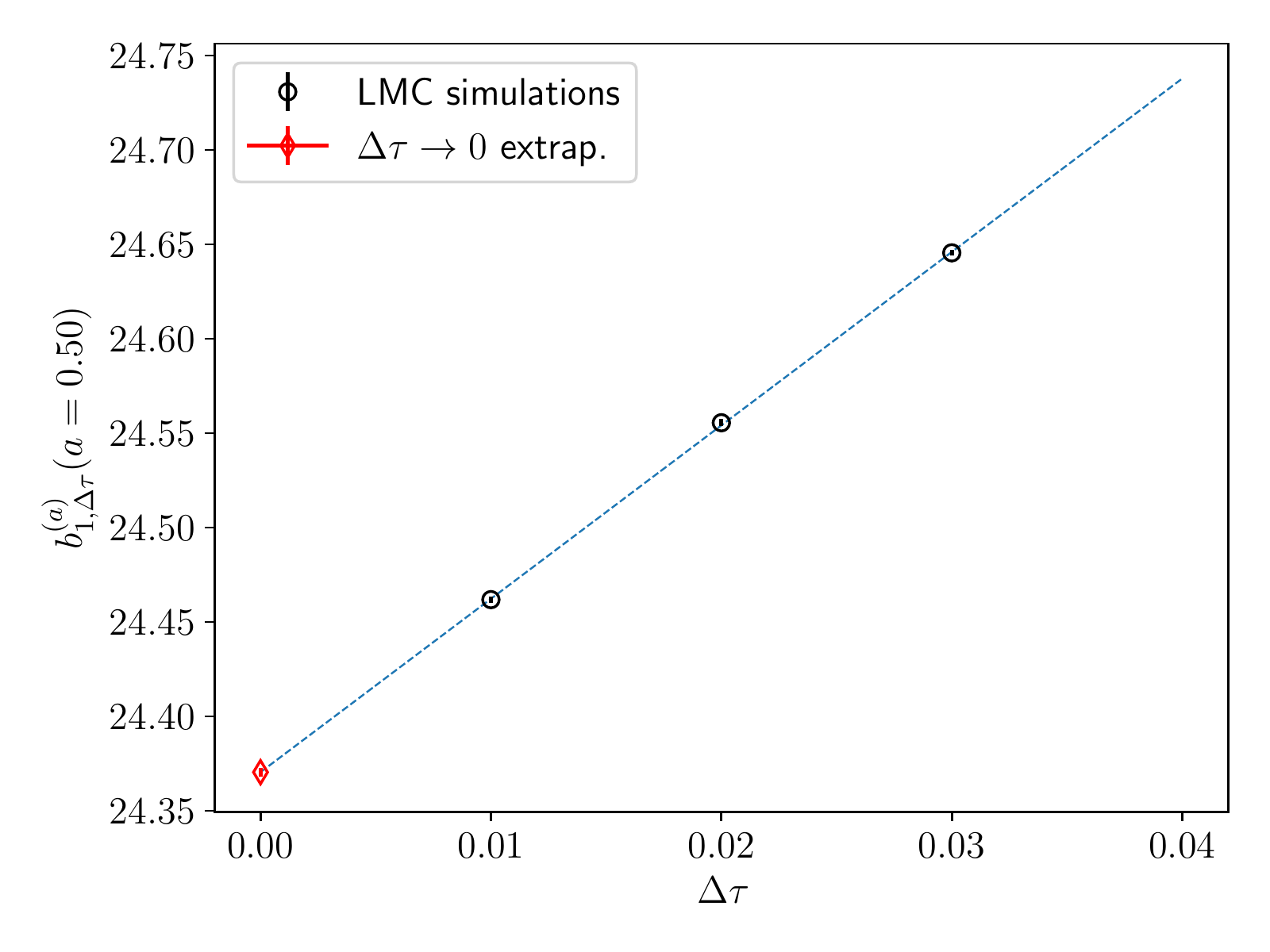}} \\
      {\includegraphics[width=.48\textwidth]{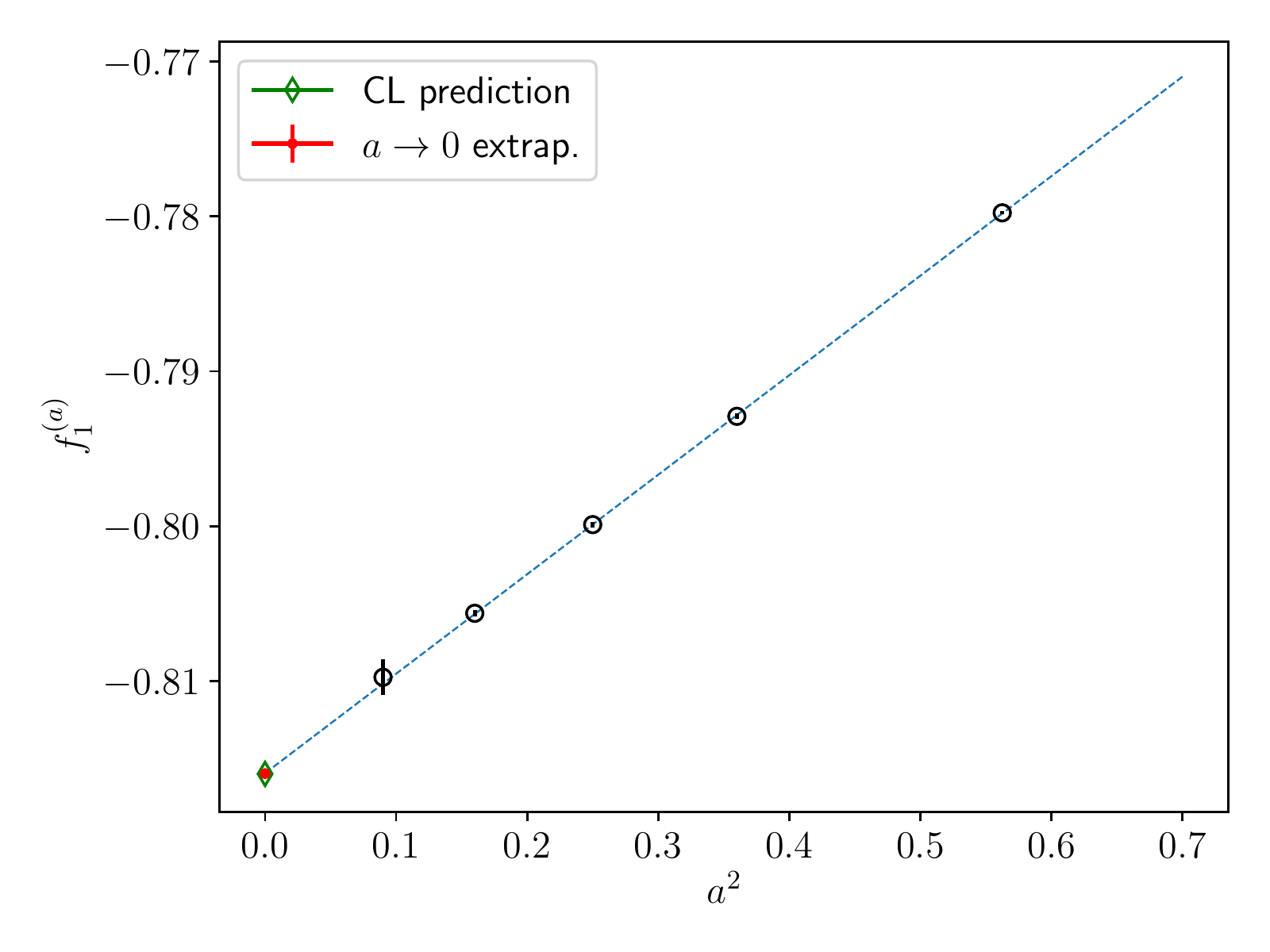}} \quad
   {\includegraphics[width=.482\textwidth]{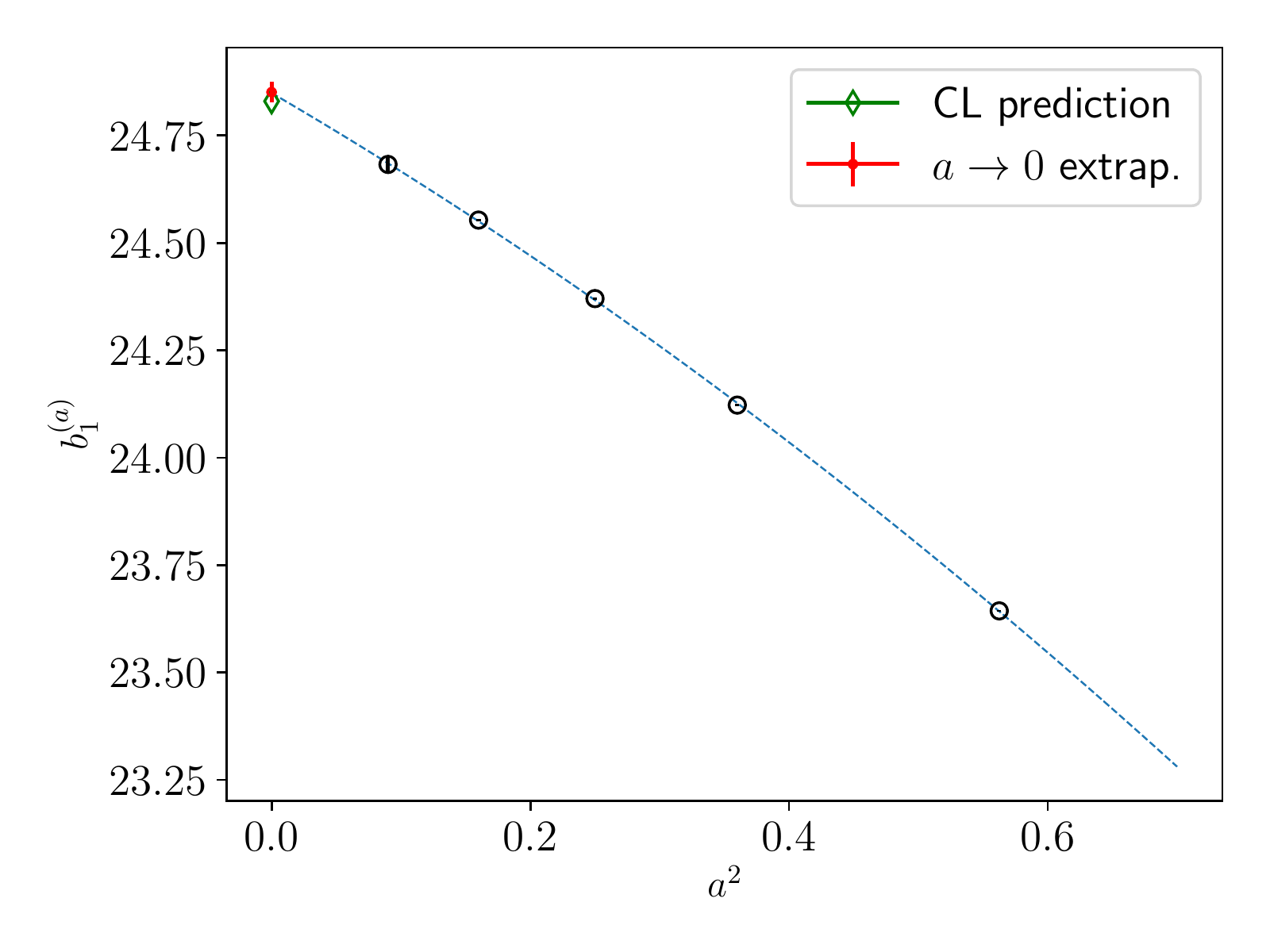}}
\caption{\textit{First row}: examples of continuum stochastic time extrapolation. We present the first-order coefficients $f_1^{(a)}$ (right) and $b_1^{(a)}$ (left). $f_1^{(a)}$ and $b_1^{(a)}$ are the dimensionfull counterpart of $\tilde{f}_1^{(a)}$ and $\tilde{b}_1^{(a)}$ defined in Eq.~(\ref{eq:f_def}) and Eq.~(\ref{eq:b_n}) and are computed at fixed values of the lattice spacing $a$ ($a=0.75$ and $a=0.50$ respectively). \textit{Second row}: continuum limit extraction for the same coefficients: $f_1$ (right) and $b_1$ (left).}
\label{fig:fig_4}
\end{figure}

Before showing some preliminary results, let us focus the leading-order expression in Eq.~(\ref{eq:Za_over_Z}) 
$$\frac{e^{-S_E[\tilde{x}^*]}\beta\sqrt{S}}{\sqrt{2\pi}}\Bigl( \frac{Z_{a,\perp}^{(0)}}{Z^{(0)}} \Bigr).$$ 
This is expressed in terms of finite lattice spacing quantities which in Fig.~ \ref{fig:fig_3} are shown to converge to their continuum limit values (we present two of them). This is an important piece of information since we will consistently extract the continuum limit of $\Delta E_0$ defined in Eq.~(\ref{eq:ration}).
It seems that the continuum limit for these quantities is well under control.\\
The perturbative coefficients we are eventually interested in are defined in the continuum limit. At any given value of the lattice spacing $a$ we have till now defined dimensionless quantities, {\em i.e.} the $\tilde{f}_n^{(a)}$ of Eq.~(\ref{eq:f_def}) and $\tilde{b}_n^{(a)}$ of Eq.~(\ref{eq:z_a_p}). Corresponding dimensionfull quantities are defined {\em e.g.} by $b^{(a)}_n = (\frac{a^5}{12})^n\tilde{b}_n^{(a)}$, where the additional factor $12$ is there to comply to the notation of \cite{escobar-ruiz_three-loop_2015}. For the same purpose of comparing to the results in \cite{escobar-ruiz_three-loop_2015} we make the choice $m=\omega=1$ (which results in different values of $\tilde{m}$ and $\tilde{\omega}$ at each values of the lattice spacing $a$). \\
In this work Langevin equation (\ref{eq:langevin}) has been integrated in the Euler scheme. 
Before we extrapolate results to the continuum limit $a \to 0$, we need to extrapolate them to the continuum stochastic time limit.
At a given perturbative order $n$ one could naively think of extracting the continuum stochastic time limit by fitting the measurements $\alpha^{(a)}_{n,\Delta \tau}$ taken at a given time step $\Delta \tau$ and lattice spacing $a$ to the expression
$$ \alpha^{(a)}_{n,\Delta\tau} =  \beta^{(a)}_{\alpha,n}\Delta\tau + \alpha^{(a)}_{n}$$
the fit being defined by the minimization of a convenient $\chi^2$. Notice that the 
$\alpha^{(a)}_{n,\Delta\tau}$ can be either the $f^{(a)}_{n,\Delta\tau}$ or the $b^{(a)}_{n,\Delta\tau}$.
Such a fit would be inconsistent: auto-correlation and cross-correlation between different orders are to be taken into account. As a result, at any given lattice spacing $a$ the continuum stochastic time limit is obtained by minimizing the quantity \cite{del_debbio_large-order_2018}
\begin{equation}
\label{eq:chi_sq}
	\chi^2 = \sum_{n,m}^{n_{max}}\sum_{\Delta\tau}\sum_{\alpha,\gamma = \{f,b\}}(\alpha_{n,\Delta\tau}^{(a)} - \beta_{\alpha,n}^{(a)}\Delta\tau - \alpha_{a}^{(n)})\mathrm{Cov}^{-1}(\alpha_n,\gamma_m)_{\Delta\tau}(\gamma_{m,\Delta\tau}^{(a)} - \beta_{\gamma,m}^{(a)}\Delta\tau - \gamma_{m}^{(a)}).
\end{equation}
For completeness we stress that such a fit is meaningful in the region where linear scaling in $\Delta\tau$ is obtained. Integrated auto-correlation and cross-correlation times are computed according to \cite{sokal_monte_1997}.\\
Examples of the two extrapolations ($\Delta \tau \to 0$ and $a \to 0$) are given in Fig.~\ref{fig:fig_4}. In the first row, we show two examples of continuum stochastic time extrapolation (for coefficients $f^{(a)}_1$ at $a=0.75$ and $b^{(a)}_1$ at $a=0.5$). In the second row we show the continuum limit extrapolation for the first-order Faddeev-Popov term $f_1$ and the coefficient $b_1$. Notice that $f_1$ and $b_1$ are now defined in the continuum limit (and thus they do not have any superscript). The NSPT predictions seem to agree with the continuum perturbative corrections, within a reasonable uncertainty. The final perturbative correction for the energy-splitting can be computed from Eq.~(\ref{eq:Za_over_Z}) and at first-order reads
\begin{equation}
\label{eq:final_coeff}
	z_1 = f_1 + (b_1 - c_1).
\end{equation}
Our preliminary result is $z_1 = -0.966(25)$ that agrees with the PI computation $z_1 = -71/72 \approx -0.986$ \cite{wohler_two-loop_1994}. Although extrapolations for $f_1$ and $b_1$ are very accurate, the final result reported here suffers from a not-too-negligible relative error because of a cancellation effect in Eq.~(\ref{eq:final_coeff}). In fact, the subtraction involves large but very similar contributions. 

\section{Conclusions and outlook}

In this work we have provided an idea of how NSPT can compute perturbative corrections around instantonic solutions. We have seen that (at least for now) the subtle challenges involved are only partially solved by NSPT. In fact, the observables we need (even in this simple system) require high-precision measurements. It is expected that difficulties may increase due to exponentiation in Eq.~(\ref{eq:Za_over_Z}) and large statistical fluctuations which can occur \cite{alfieri_understanding_2000} at high-orders.
Actually, at the conference we had the chance to talk to another group that has been tackling the same computations; they told us that large fluctuations at high-order indeed occur \cite{priv_comm}, an effect that we found ourselves after the conference. Certainly we know from previous experience that life can be actually easier for larger systems, thus opening the path to a more successful application of the method in quantum field theory. 

\section{Acknowledgments}
We thank Alberto Ramos and Guilherme Catumba for very interesting discussions. This work was supported by the European Union Horizon 2020 research and innovation programme under the Marie Sklodowska-Curie grant agreement No 813942 (EuroPLEx) and by the I.N.F.N. under the research project (\textit{iniziativa specifica}) QCDLAT. This research benefits from the HPC (High Performance Computing) facility of the University of Parma, Italy.

\bibliographystyle{JHEP}
\bibliography{my_bib}

\end{document}